\documentclass{article}
\begin{document}

\begin{title}
\centerline{\large \bf The role of intense line
 wings in the process of infrared
multiple--photon excitation of polyatomic molecules.}
\end{title}

\vspace{3 pt}
\centerline{\sl V.A.Kuz'menko}
\vspace{5 pt}
\centerline{\small \it Troitsk Institute for Innovation and Fusion Research,}
\centerline{\small \it Troitsk, Moscow region, 142190, Russian Federation.}
\vspace{5 pt}
\begin{abstract}

 The results of experimental testing  the existence of intense
 Lorentzian--like wings with FWHM $\sim 4.5 cm^{-1}$ in
 the absorption spectra of polyatomic molecules in a gas phase are presented.
 Two independent experimental methods were used for evaluating the integral
 intensity of the line wings for a number of substances.  In the first case,
 the cross--section of the far wings of absorption bands in a gas phase
 spectrum were  measured.  Then, these band wings were extrapolated inside the
 contour of absorption band.  In the second case, the saturation degree
 of the linear spectrum of molecules  was determined. Radiation of a pulsed
 $CO_2$--laser was used  at low gas pressure ( $\sim 16$ mtorr) and averaged
 excitation level  of molecules ${<n>}\sim 0.1$ quanta/molecule.  The values
 obtained by these two independent methods coincide for a variety of molecules.
 The average relative integral intensity of the line wings varied
 from $\sim 0.6\%$  for  $SF_6$ and $SiF_4$  to $\sim 90\%$ for  $(CF_3)_2O$
 and  $(CF_3)_2CO$. 

\vspace{5 pt}
{PACS number: 33.70.-w}
\end{abstract}

\vspace{12 pt}

\section{Introduction}

 The phenomenon of the infrared multiple--photon excitation (IR MPE) and
 collisionless dissociation of polyatomic molecules was discovered
 in works [1,2]. 
 Numerous works were carried out later aimed to clearify the 
 mechanism of this process.  This interest was stimulated by the fact, 
 that the widths both of laser radiation and molecule absorption lines 
 are several orders of magnitude lower, than the anharmonicity  of
 molecular vibrations.
 In the first work [1] the hypothesis of the existence of the
 so--called "quasicontinuum" of vibrational states was proposed. 
 Despite some argued criticism [3], this idea has got the broadest 
 distribution.

 In the recent years, however, the views on the nature of 
 "quasicontinuum" have changed dramatically. 
 Earlier, it was accepted, that "quasicontinuum" consists 
 of a huge number of narrow lines arising as a result of coupling 
 different vibrational states.  Now it is widely believed [4], 
 that the absorption line is unique, but it becomes very wide. 
 The origin of the "quasicontinuum" now is bounded up with 
 intramolecular vibrational relaxation (IVR). 
 This is a reasonable idea. 
 The IVR process can be very fast (picosecond timescale). 
 The corresponding Lorentzian width of the absorption line can be
 in this case comparable with anharmonisity of the molecular vibrations.
 The main disadvantage of this model is that it does not explain how 
 the molecules can be excited in the field of low vibrational levels,
 where the IVR is absent and the absorption lines remain narrow. 
 Experiments show, that excitation of molecules in this area occurs
 without essential difficulties, but the theory gives no satisfactory
 explanation of this fact.

 In works [5,6] an idea was proposed, that the IR MPE process 
 is a trivial result of absorption in the area of line wings, but
 untrivial is the nature of these wings. 
 Practically, the possible role of line wings was not discussed
 in the literature earlier. 
 It is, apparently, due to the fact, that appropriate estimations 
 can easily be made.  The lifetime of excited states of molecules
 due to spontaneous emission in the infrared region lays in the millisecond
 timescale.  The natural width of line must to be smaller than $100 Hz$. 
 Even for the strongest molecular transitions, at the
 distance from the line center equal to the value of molecular angarmonicity,
 the Lorentzian countour of the natural width 
 would have an absorption cross--section smaller then $\ 10^{-25}\ cm^2$.
 This cross--section cannot play any appreciable role in
 overcoming the angarmonicity of molecular vibrations.

 However, such an estimation has not been tested in experiment earlier.
 It is possible to assume, that for some unknown reasons, 
 intensity of real line wings is much higher, than the theory predicts. 
 How high the intensity of line wings should be to explaine 
 the observable effect of laser excitation of molecules? 
 Rather correctly such information can be derived from 
 the experimental results of works [7-9], where the depletion of rotational
 states of  $SF_6$ molecules by TEA $CO_2$--laser radiation was studied
 in the conditions of molecular jet. 
 The results of such processing are presented in Fig.1.
 Except for the usual narrow component of the line with a Doppler width 
 $\sim 30 MHz$,  the wings, or more precisely speaking, 
 a wide component of the line should exist with a
 cross--section $\sigma\simeq 6\cdot{10^{-20}} \ cm^2$ and with a Lorentzian
 full width at half medium $2 \Gamma\sim 4.5\ cm^{-1}$. 
 The relative integral intensity of this component is rather small, 
$X\sim 0.2\% $, but it is high enough for efficient excitation of molecules 
from all rotational states.

 The purpose of the present work is an experimental testing the hypothesis
 of the existence of intense line wings.  First of all, it is important
 to point out that the Lorentzian countor is rather flat and the wings of
 the line can manifest themselves as far natural wings of absorption bands.
 Numerous literature exists about the far wings of absorption 
 bands for small and light molecules which are due to collisional broadening
 of the lines [10,11], but publications devoted to the far wings of absorption
 bands in the gas phase for heavy polyatomic molecules are absent.  In the 
present work, for the first time, far natural wings of absorption bands are 
discovered for a number of polyatomic molecules. The experiments show, that 
the absorption cross--section in the region of these far band wings is 
independent on the gas pressure. Measuring the intensity of the far band 
wings allows to estimate the relative integral intensity of the line wings.

 On the other hand, it is important to pay attention to the phenomenon of
 saturation of the  absorption spectrum of polyatomic molecules by laser
 radiation at low gas pressure [12]. It is well known, that for some molecules 
this saturation can be very pronounced, but for the others it is minimal [13]. 
 A satisfactory explanation of this distinction was not found. 
 From the point of view of the discussed hypothesis, the distinction 
 is due to the difference in relative intensities of the line wings: 
 for the molecules with a large relative intensity of
 the line wings the saturation is minimal and vise versa.

 So, the saturation degree of the absorption spectrum of polyatomic 
 molecules can be used to estimate the relative integral intensity 
 of the line wings. With this purpose, in the presented work,
 the saturation degree of the linear absorption spectra of a number of
 molecules by  the TEA $CO_2$--laser radiation  was investigated.
 The measurements were conducted in comparable conditions 
 at low gas pressure and at average level of molecular excitation
 $<n>\sim 0.1\ $ quanta/molecule. 
 It was supposed, that under these conditions the narrow component of the line 
 is completely saturated, but the wide one is not.
 Then the measured saturation degree can be regarded as a rough
 estimation of the relative integral intensity of the line wings.
 Estimations of the line wings intensity obtained by two 
 independent methods were compared with each other for a number of molecules.

\section{ Experimental}

The TEA $CO_2$--laser with an additional low pressure cell
inside the resonator was used.
When a low pressure cell was switched on, the TEA $CO_2$--laser 
switched from the multimode regime to the single--mode one.
Using a nitrogen free gas mixture allowed to obtain a pulse of generation
with $\sim 150 ns$ duration at half height without a "tail".

A piroelectric detector was used for measuring the laser
radiation absopbtion at low gas pressure. 
The technique of experiments is similar to that described in [14]. 
The piroelectric detector was placed inside the cell
(pass lenght $110 cm$) on the distance $10 mm$ from the laser beam. 
Calibration of the detector's sensitivity was carried out by measuring
the attenuation of the transmitted beam at relatively high gas pressure
($>100 mTorr$).

Measurements of the absorption cross--section in the region of the far 
band wings in some cases were carried out with an optico--acoustic 
detector (OAD).  This detector was also used to study the influence of the 
buffer gas pressure on the intensity of the far band wings. 
In the last case, first of all, the dependence of the OAD sensitivity 
on the gas pressure was measured. 
The cell was filled with $\sim 40$ mTorr of hexafluoroacetone and it
was irradiated by TEA $CO_2$ laser on 10R(12) line with energy $\sim 1mj$.
Under these conditions, the amount of energy, absorbed in the cell, remains 
constant, and the dependence of the OAD signal on the buffer gas (nitrogen) 
pressure, becomes the dependence of the detector's sensitivity on the 
gas pressure. Then the cell was filled with $1 Torr$ of the substance to study 
($CF_2Cl_2, SiF_4$) and $1-800 Torr$ of nitrogen was added. 
The energy of a pulse laser radiation was usually $30-70 mj$ 
and it's wavelenghth was selected in the region, where
the far band wings had sufficiently clean countor and the absorption
cross--section in the range $10^{-21}-10^{-22} cm^2$.

An IR--spectrophotometer with a stainless steel cell (pathlenth $10 cm$ and gas
pressure up to $5 atm$) was used to study the absorption band profile for a
number of molecules. Most of the used substances were commertially 
available and had been used without purification. Completely fluoreneted
dimethylether was obtained by rectification of the products' mixture after
fluorenation of the dimethilether. The gas used in the experiments contained
$ \sim 97 \% $ of the basic substance.  As an impurity, $(\sim 3\% )$ of the
 $CF_3H$ was present.

\section{ Results}

  Typical results of the experimental study are presented in Figs. 2 and 3.
These pictures show the saturation degree of linear
absorption spectrum by pulse laser radiation of the $SF_6$ and $(CF_3)_2CO$ 
molecules. In the first case the saturation is very high, and in the second
 one it is minimal in full agreement with early results [13,15]. 
At average level of excitation $\sim 0.1 $ quanta/molecule, with 1 atm of
 buffer gas added, the measured absorption cross-section of the laser radiation 
practically coincides with the same cross-section ($\sigma_0$), obtained with 
a spectrophotometer.

Table 1 contains main results of these experiments. 

\begin{center}

Table 1.{\bf Determination of the saturation degree of 
the linear absorption spectrum of polyatomic molecules by pulse 
$CO_2$--laser radiation.} 

\vspace{12pt}

\begin{tabular}{|c|l|l|c|c|c|c|c|}                       \hline
{\bf N} & {\bf Molecule} & {\bf Laser} & {\bf Fluence} & {\bf Pressure} & $\sigma_m / \sigma_1$
  & $<n>^*$    	& $\sigma_1/\sigma_0$                   \\ 
  & 	     	& {\bf line}   & $mj/cm^2$ & $mTorr$ & & & \%  \\ \hline
1 & $SF_6$ 	& 10P(20) & 32 	& 16 	& 2.0 	& 0.15 	& 0.6  \\
  &      	&         & 32 	& 100 	& 1.8 	& 0.25 	& 1.0  \\ \hline
2 & $SiF_4$ 	& 9P(34)  & 32 	& 100 	& 1.4 	& 0.07 	& 1.0  \\ \hline
3 & $CF_2Cl_2$ 	& 10P(36) & 32 	& 100 	& 1.3 	& 0.06 	& 1.5  \\ \hline
4 & $CFCl_3$ 	& 9R(30)  & 32 	& 100 	& 1.4 	& 0.07 	& 4.0  \\ \hline
5 & $C_2F_3Cl$ 	& 9R(6)   & 32 	& 100 	& 1.4 	& 0.15 	& 12   \\ \hline
  &          	&         & 32 	& 16 	& 1.3 	& 0.35 	& 7.0  \\    
6 & $CH_3SiF_3$	& 10R(26) & 32 	& 100 	& 1.3 	& 0.55 	& 11   \\
  &           	&         & 3.2 & 100 	& 1.4 	& 0.09 	& 17.5 \\ \hline
7 & $(CF_3)_2O$	& 10R(20) & 32 	& 16 	& 1.05	& 1.0 	& 80   \\
  &           	&         & 3.2 & 100 	& 1.05 	& 0.1 	& 90   \\ \hline
8 & $(CF_3)_2CO$ & 10R(14) & 32 	& 16 	& 1.00 	& 2.8 	& 70 \\
  &            	&         & 3.2 & 100 	& 1.00 	& 0.35 	& 90   \\ \hline
9 & $C_6H_5SiF_3$ & 10R(10) & 32   & 16  & 1.00  & 3.0 	& $\ge$ 90 \\
  &             &         & 3.2 & 100 	& 1.00 	& 0.3 	& $\ge$ 90 \\ \hline
\end{tabular}
\end{center}
\vspace{8pt}
${}^* <n>=\sigma_1 F/h\nu$

\vspace{12pt}

The measured absorption cross-sections are usually greater 
for multimode laser radiation ($\sigma_m$) 
than for the singlemode one ($\sigma_1$) [16]. Owing to limited
sensitivity of the piroelectric detector the minimum used gas pressure
was chosen 100 mTorr in some cases. For these cases, the obtained final value
$X={\sigma_1}/{\sigma_0}$ is overestimated by a factor of $\sim 1.6$.

In Fig.4, the spectral dependence of the absorption cross-section of $SiF_4$
molecules around the $\nu_3$ absorption bands is presented. 
The edges of the absorption band have approximately an exponential form, 
the slope beeng greater for the blue side, than for the red one. 
At the distance more, than 25--40 $cm^{-1}$ from the band center, much more 
glantly sloping wings are observed. The curve (2) is a Loretzian profile with
 FWHM $2\Gamma=4.5 cm^{-1}$, which passes through the point with minimal
 absorption cross-section in the given spectral range. 
So, we can see, that far band wings have a Lorentzian behavior.

In Fig.5 a spectral dependence of the absorption cross-section of $SF_6$ 
molecules is shown. In this rather typical case the far band wings are masked
 by intense combination bands.

The same results for the $CF_2Cl_2$ molecules are shown in Fig. 6. In the
 latter case the far band wings can be observed in rather limited spectral 
range $950 - 1000 cm^{-1}$.

Fig.7 shows the spectrum of the $(CF_3)_20$ molecules. 
The basis of the spectral dependence well coordinate with a Lorentzian
 contour, which center coincides with the most intense absorption band. 
It can be seen also, that intensity of the Lorentzian contour in this case 
approximately two orders of magnitude higher, than for the cases 
of $SiF_4$ and $SF_6$ molecules, although the intensities of the strongest
 absorption bands are comparable for all these molecules.

The experiments have shown, that the absorption cross-section of the
investigated molecules in the region of far wings of their absorption bands
depends neither on it's own gas pressure, nor on the buffer gas pressure
(in the range 8-- 800 Torr of nitrogen). It means, that the far wings
of absorption bands are natural and have no relation to the collisional
broadening of the lines.

The computed relative integral intensities of the line wings are listed 
in Table 2. It was supposed, that $2\Gamma=4.5 cm^{-1}$ in all cases.
For comparison, the final results of Table 1 are also shown here. 
It is visible,that the difference in relative integral intensities of the 
line wings reaches two orders of magnitude for different molecules. 
At the same time, the maximum deviation between the estimations, 
obtained by two independent methods, does not exceed three times. 
The intensity of the line wings is rather small for the 
molecules with one central atom and grows quickly with increasing 
the number of atoms and branching degree of the molecules.

\newpage

\begin{center}

Table 2.{\bf Evaluation of the relative integral intensity of the line wings.}

\vspace{10pt}
\begin{tabular}{|c|l|c|c|}                                  
\hline
N & Molecule & \multicolumn{2}{|c|}{ X \%}   	      \\ 
 	       \cline{3-4}				  
  &          	& {\bf From the band wings} & {\bf From the saturation degree}
 \\ \hline
1 & $SiF_4$ 	& 0.6 		& 0.6                 \\ \hline
2 & $SF_6$  	& 0.8 		& 0.6                  \\ \hline
3 & $CF_2Cl_2$ 	& 1.6 		& 0.9                  \\ \hline
4 & $CFCl_3$ 	& 2.3 		& 2.5                  \\ \hline
5 & $C_2F_3Cl$ 	& 3.0 		& 7.5                  \\ \hline
6 & $CH_3SiF_3$	& 11 		& 11                   \\ \hline
7 & $(CF_3)_2O$	& 36 		& 90                   \\ \hline
8 & $(CF_3)_2CO$ & 40 		& 90                   \\ \hline
9 & $C_6H_5SiF_3$ & 50 		& $\ge$ 90             \\ \hline
\end{tabular}
\end {center}

\section{ Discussion}

Close, although not so precise as from works [7-9], estimations 
of the parameters of the wide component of line for $SF_6$ molecules 
($\sigma\sim 10^{-19}cm^2$  and  $ \Gamma<3 cm^{-1}$) can be obtained by processing 
the experimental results of work [17]. 
In this latter work the absorption of CW $CO_2$ -laser radiation by
$SF_6$ molecules was studied in the molecular beam conditions, 
and the autors dealt practically only with the line wings.

A substantial difference exists between the estimations 
of the relative integral intensity of the line wings for $SF_6$ molecules, 
derived from works [7-9] $(X\simeq 0.2\% )$ and obtained 
in the present work $(X = 0.6 - 0.8\% )$. This is obviously due to the fact,
 that in the first case the molecules were located at a zero vibrational level,
and the second estimation is obtained for the room temperature, 
when the majority of molecules is in different excited states. 
The strong temperature dependence of the laser radiation absorption by
$SF_6$ molecules [17] allows to assume, that the intensity of line wings
substantially grows with increasing the level of vibrational excitation of
molecules. This is maybe the main physical reason of formation of ensembles
of "hot" and "cold" molecules under action of the pulse $CO_2$ laser
radiation [18].

It is visible from Table 2, that for the largest molecules a systematic
underestimation exists when the intensity of the far band wings is used.
This is maybe due to the real width of line wings for large molecules
is smaller, than it was supposed in calculations. 
Such an assumption agrees with the results of experimental study of 
photodissociation of a number of molecular dimers in [19], where the  
the autors dealt generally with the same wings of lines. 
With increasing the molecular weight the spectral photodissociation width
of the corresponding dimer usually decreases: for
$(C_2H_4)_2), (SiF_4)_2, and (SF_6)_2$ the corresponding values are 12, 4.7, 
and $1.5 cm^{-1}$, accordingly.

The difference in absorption of the singlemode and multimode laser radiation
has, obviously, common nature with a similar effect, observed earlier in works
[16,17]. This is possible only in the conditions, when the IVR process is
 absent in a molecule. Therefore, it can be expected, that in the $(CF_3)_2CO$
 and $C_6H_5SiF_3$ molecules the fast IVR occurs already after the absorption 
of the first quantum of $CO_2$ laser radiation. More precisely this question
 could be clearified after studying the full spectrum of these molecules in 
the conditions of a free jet, when the IVR process can frequently be 
identified [20].

The results of the present work, combined with experimental data of works
[7-9, 17],  provide quite sufficient proof of existence of the intense
line wings in polyatomic molecules. The nature of these wings can be related 
with breaking the mechanism of averaging the molecular inertia moment on the 
timescale of vibrational motion.
The spectroscopic methods allow a high precision determination 
of the inertia moment of a molecule, this moment being an averaged inertia
moment because it changes by $5 - 10 \% $ on the timescale of vibrational 
motion [21]. If one supposes that the mechanism of averaging the
inertia moment can suffer periodic short--time breakings (destructions) by
various vibrational modes, then we should get a clump of lines in the spectrum 
instead of a single narrow line. The width of this clump of lines must 
apparently be determined by the maximum amplitude of the molecule inertia 
moment variations due to vibrations. 
Such the objects have been intensively studied in the last years, 
and the autors tried to explain their nature by coupling of different 
vibrational modes [22]. The very moment of breaking-up (destruction) can be 
considered as some short--life "stressed" state of the molecule, which should 
be corresponded with some wide component in the line spectrum. 
So, while it is not clear, which effect is responsible for the spectral width 
of such a component, it is worth noting, that the width itself, retrieved
from the experiments, is comparable to the period of vibrational motion of 
atoms in a molecule.

 Molecular beam with a cryogenic bolometer is ideally suitable for
 the study of line wings.  Narrow and powerful laser radiation interacts
 mainly with the line wings [17]. Low rotational temperature in the beam 
results in that the width of  the absorption band is comparable with or less 
than that of the line wings. Therefore, for precise determination of the width 
of the line wings  it is  not necessary to investigate very far wings of the 
absorption bands.  The sensitivity of the apparatus can easily be
 calibrated using large polyatomic molecules like as hexafluoroacetone.
 Upon calibration, such an instrument can be used for measuring the intensity 
and the width of the line wings for most  polyatomic  molecules.
 Using one $CO_2$--laser, we can measure the line wing parameters 
 for the $1\leftarrow 0$ transition.  Having two $CO_2$--lasers, we can 
obtain the same information  for the $2\leftarrow 1$ transition and even for 
the $3\leftarrow 2$ one ( if at the first step we make use of one of the 
twoquantum transitions, studied in [23] ).

\section { Conclusion }

	The discussed experimental results quite clearly show the existence of
 intense line wings of unknown nature in polyatomic molecules. This line 
wings are interesting in itself. But it is also very important, that such 
kind of experiments [17] gives for the present day the only straightforward 
experimental proof of the strong time invariance violation in a 
photon-molecule interaction [24,25].

So, the task of experimental study of the line wings of polyatomic molecules 
has a grate importance. Unfortunately, the experimentalists are afraid to 
work with this interesting and unusual physical object.

\vspace{15 pt}

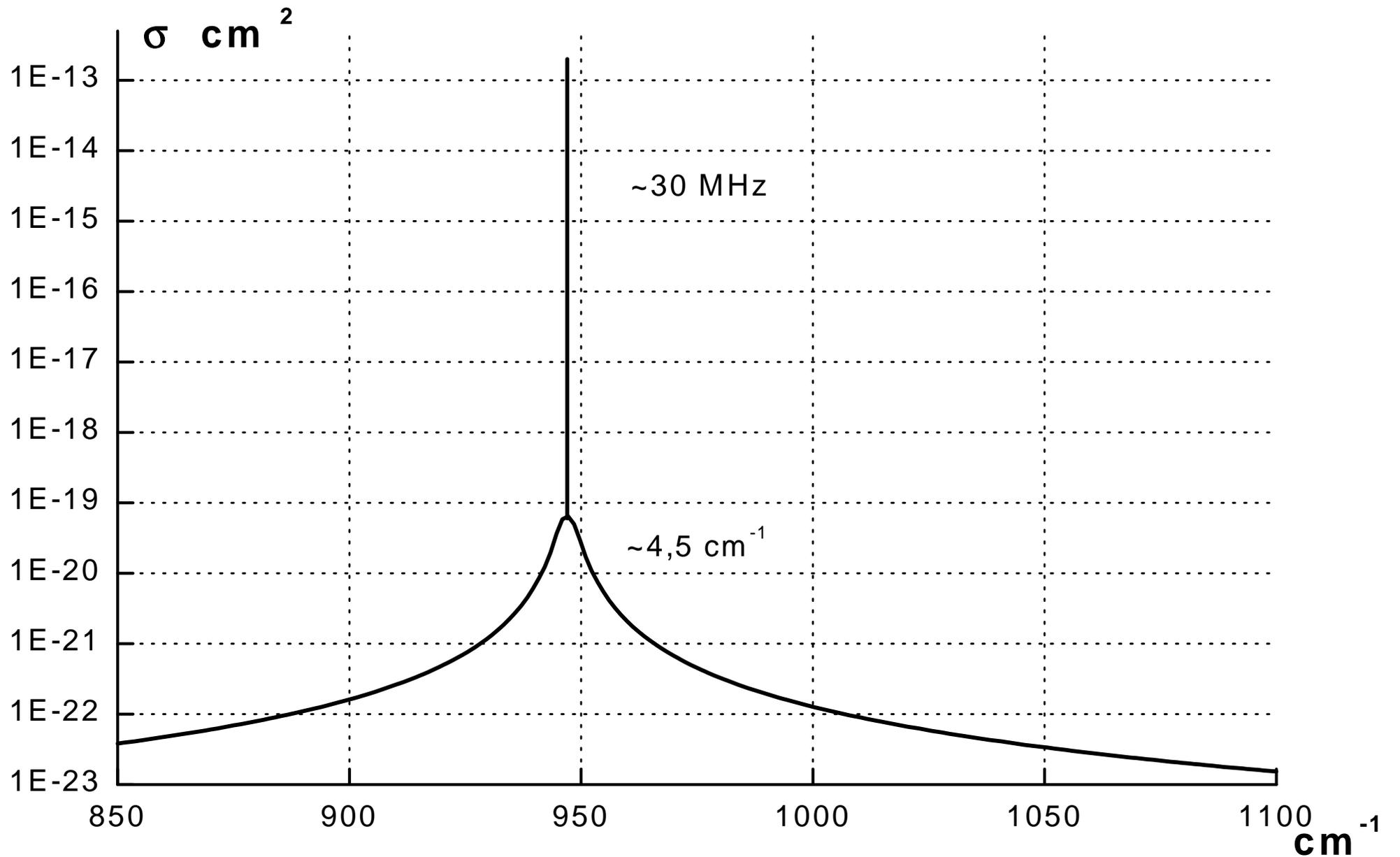
Fig.1 Profile of absorption line of $SF_6$ molecules for the $\nu_3$ band 
$1\leftarrow0$ transition.

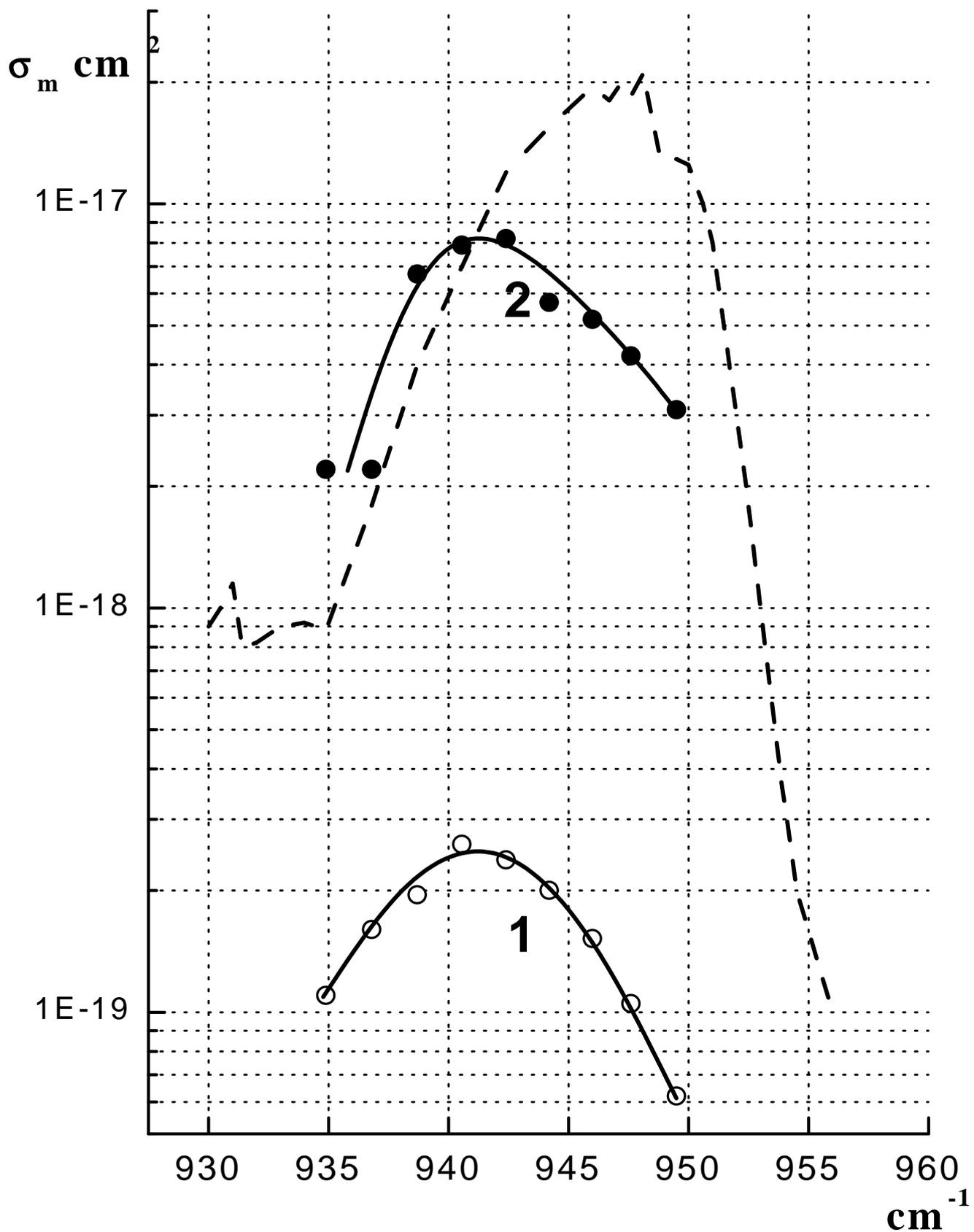
Fig.2 Spectral dependence of the absorption cross-section of multimode laser
radiation by $SF_6$ molecules. $32 mj/cm^2$.

$. -- 16 mTorr SF_6$.

$0 -- 16 mTorr SF_6 + 760 Torr N_2$.

Linear room temperature absorption spectrum of $SF_6$ molecules is also shown.

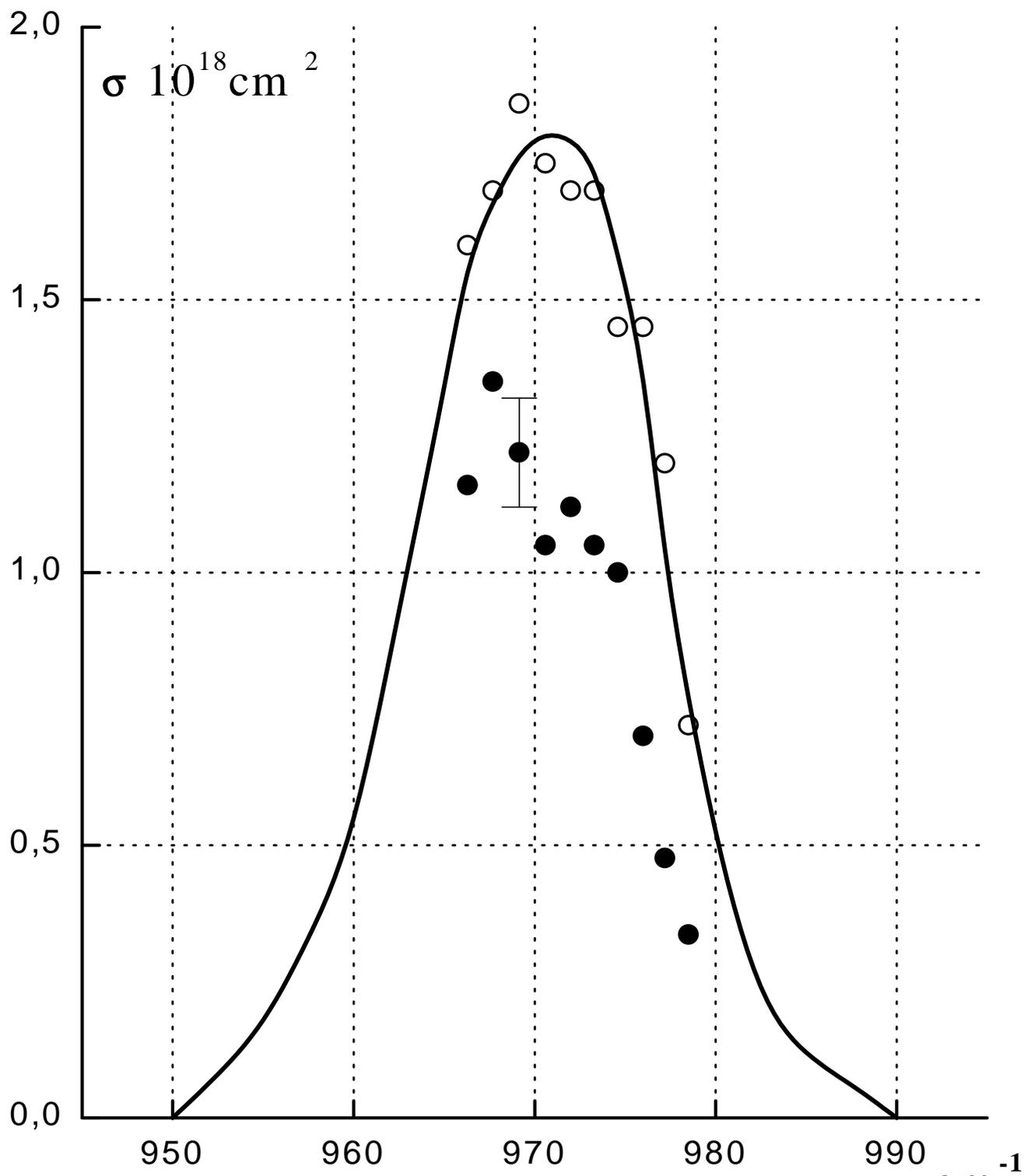
Fig.3 Spectral dependence of the absorption cross-section of multimode laser
radiation by $(CF_3)_2CO$ molecules. $32 mj/cm^2$.

$. -- 16 mTorr  (CF_3)_2CO$.

$0 -- 100 mTorr  (CF_3)_2CO + 760 Torr N_2$.

The solid curve represent a linear room temperature absorption spectrum of
 $(CF_3)_2CO$ molecules.

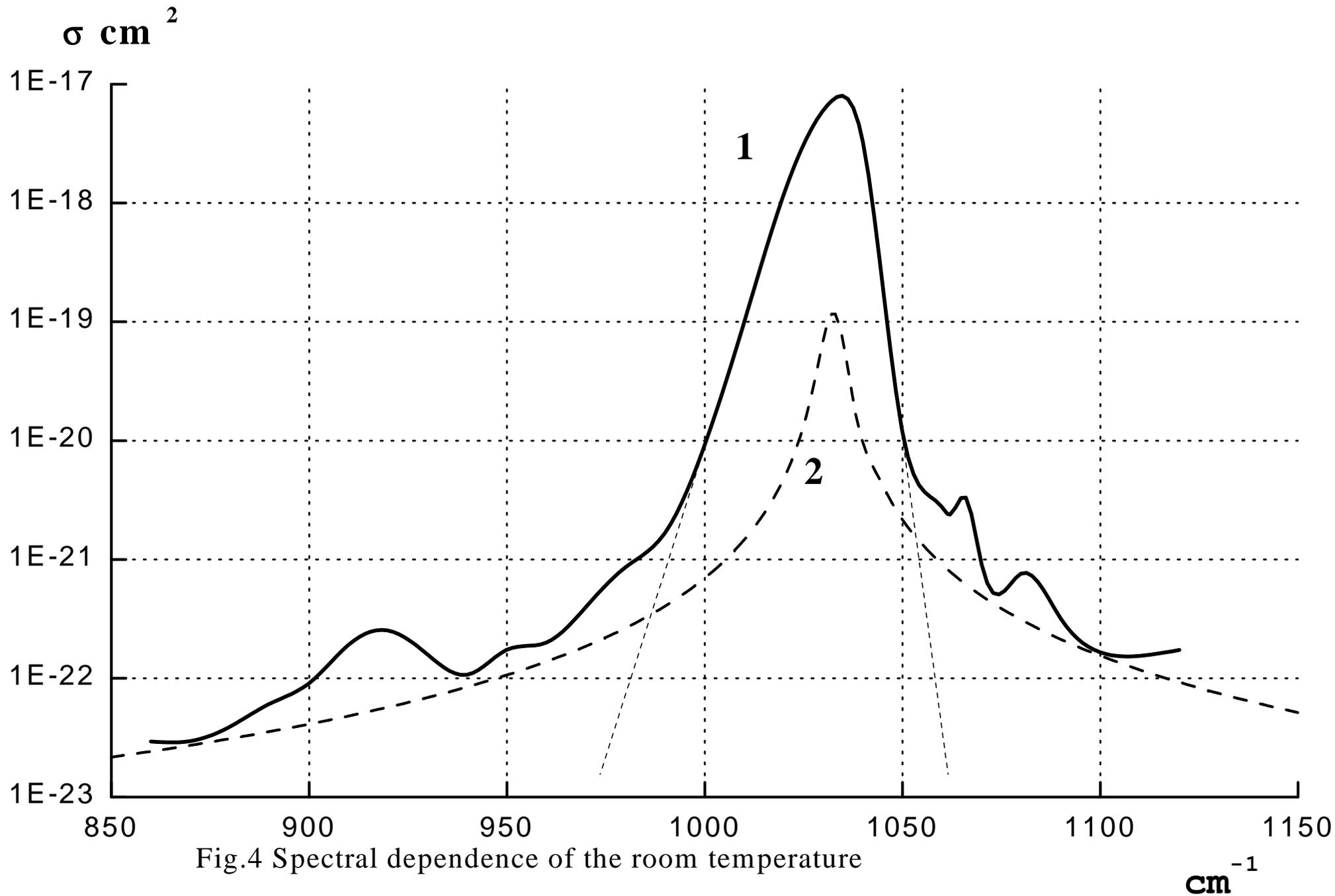
 Fig.4 Spectral dependence of the room temperature absorption cross-section
 of $SiF_4$ molecules.

1- spectrophotometer result.

2- Lorentzian profile with FWHM = $4.5\ cm^{-1}$.

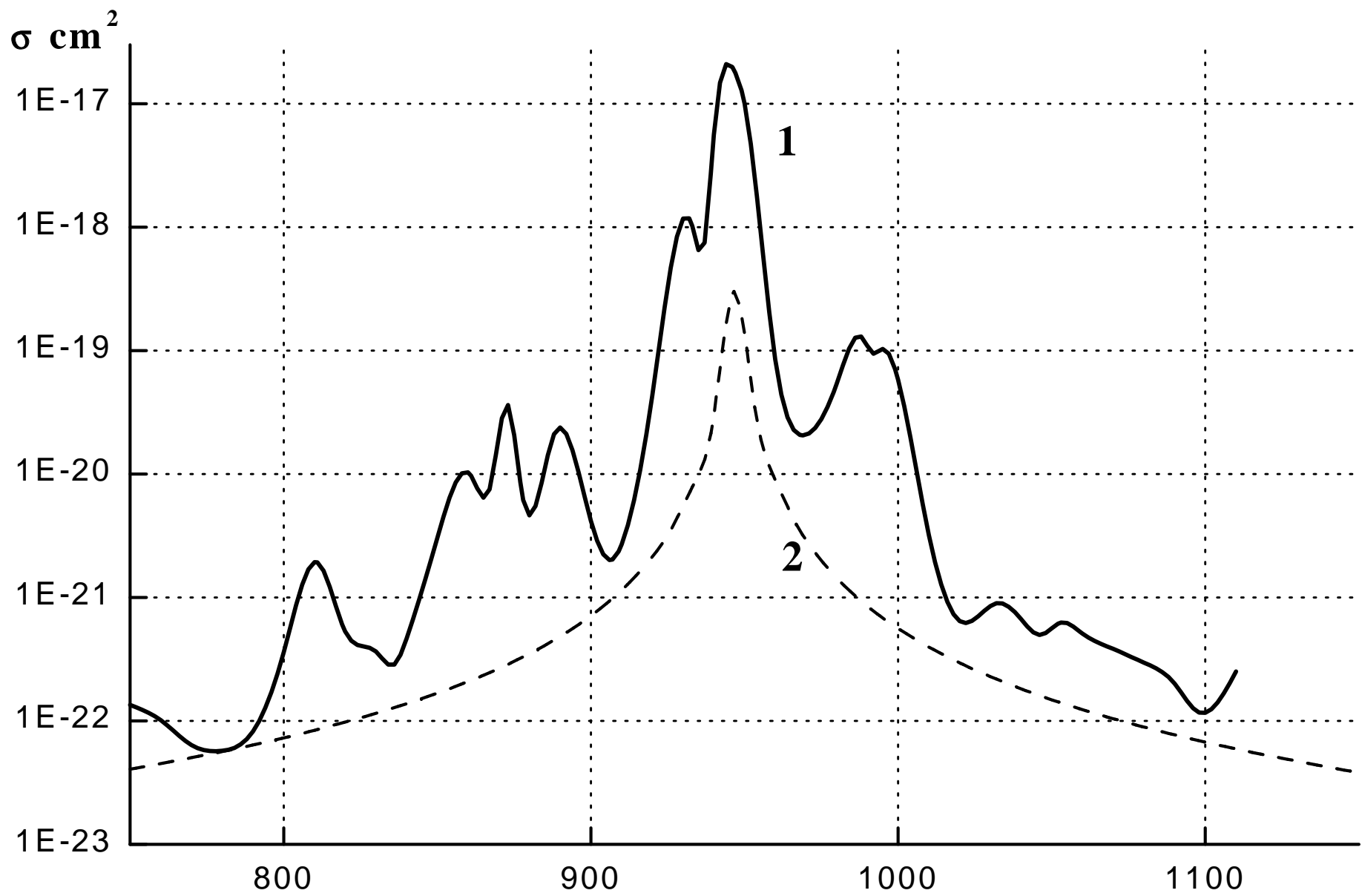
Fig.5  Spectral dependence of the room temperature absorption cross--section
 of $SF_6$ molecules.

1- spectrophotometer result.

2- Lorentzian profile with FWHM = $4.5\ cm^{-1}$.

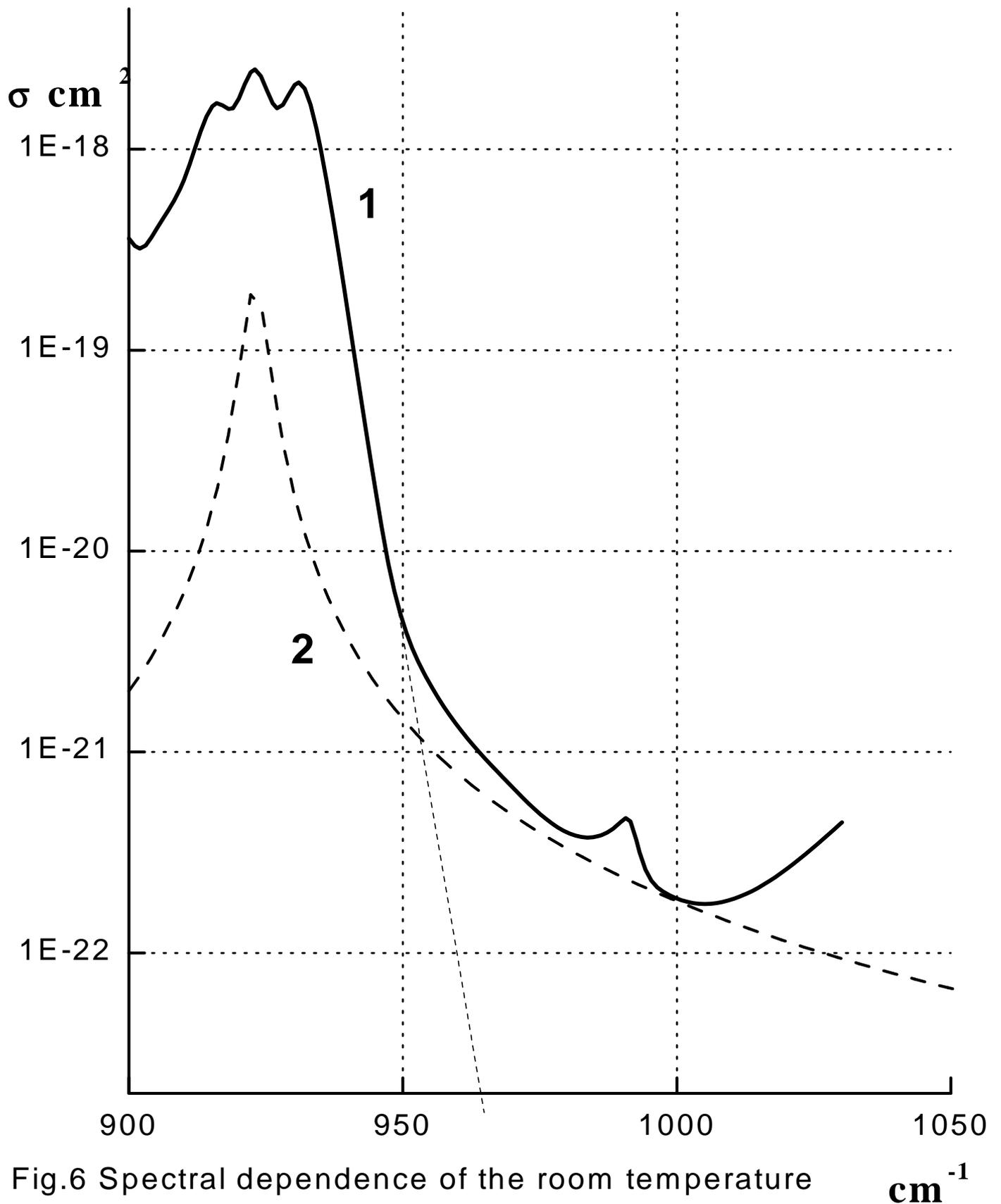
Fig.6  Spectral dependence of the room temperature absorption cross--section
 of $CF_2Cl_2$ molecules.

1- spectrophotometer result.

2- Lorentzian profile with FWHM = $4.5\ cm^{-1}$.

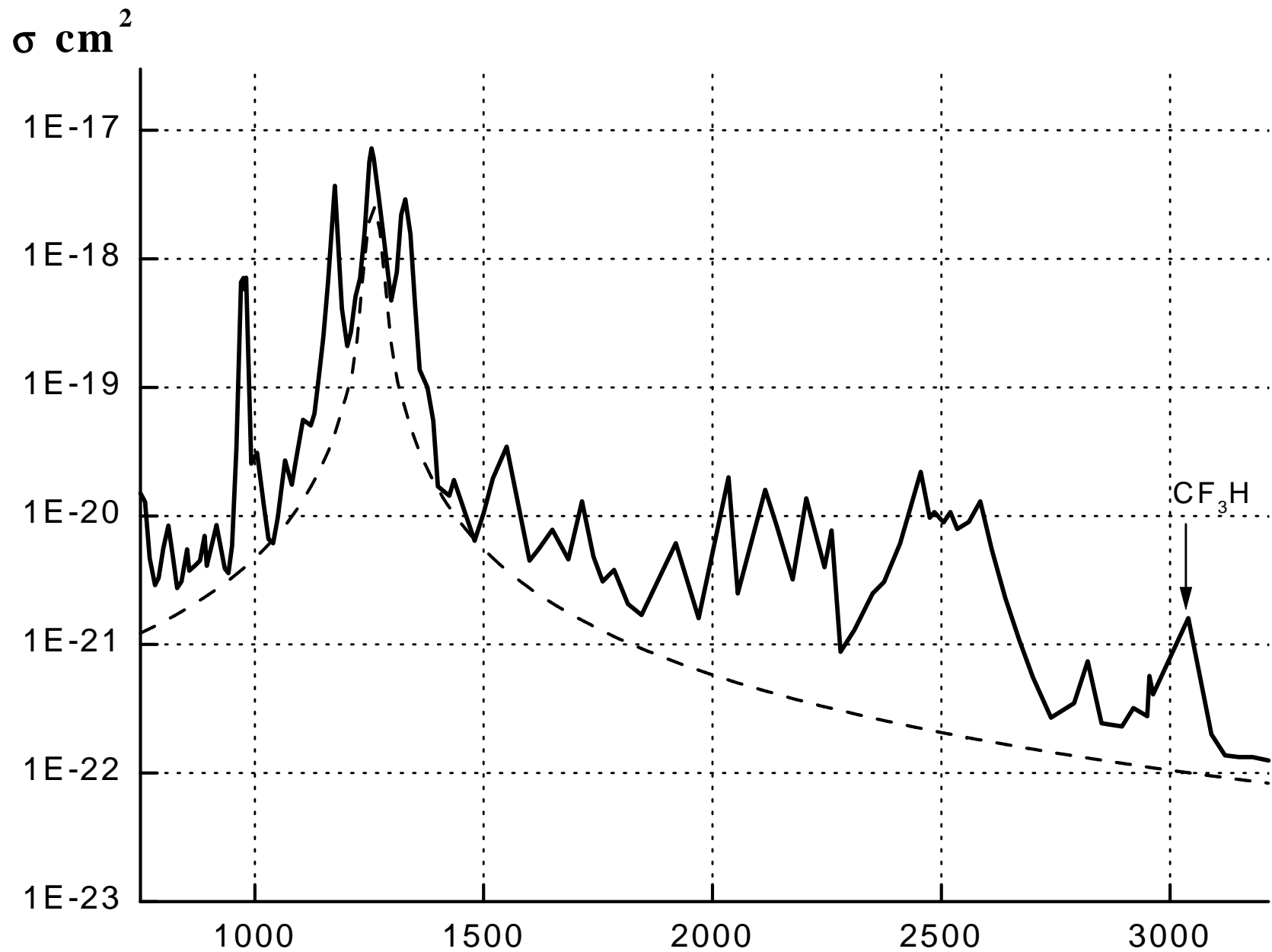
Fig.7  Spectral dependence of the room temperature absorption cross--section
 of $(CF_3)_2O$ molecules.

1- spectrophotometer result.

2- Lorentzian profile with FWHM = $4.5\ cm^{-1}$.

\end{document}